\documentclass[twocolumn]{aastex631}

\newcommand{\TESS}{{\em TESS}}
\newcommand{\Kepler}{{\em Kepler}}
\received{November 8, 2021}
\revised{December 6, 2021}
\accepted{December 13, 2021}

\submitjournal{ApJL}

\shorttitle{HD\,135348: Could it have an RRM?}
\shortauthors{Jayaraman et al.}

\begin{document}

\title{Could the Magnetic Star HD\,135348 Possess a Rigidly Rotating Magnetosphere?}
\correspondingauthor{Rahul Jayaraman}
\email{rjayaram@mit.edu}

\author[0000-0002-7778-3117]{Rahul Jayaraman}
\affiliation{MIT Kavli Institute and Department of Physics, 77 Massachusetts Avenue, Cambridge, MA 02139}

\author[0000-0003-0153-359X]{Swetlana Hubrig}
\affiliation{Leibniz-Institut f\"ur Astrophysik Potsdam (AIP), An der Sternwarte~16, 14482~Potsdam, Germany}

\author[0000-0003-2002-896X]{Daniel L. Holdsworth}
\affiliation{Jeremiah Horrocks Institute, University of Central Lancashire, Preston PR1 2HE, UK}

\author[0000-0002-5379-1286]{Markus Sch\"oller}
\affiliation{European Southern Observatory, Karl-Schwarzschild-Str.~2, 85748~Garching, Germany}

\author[0000-0003-3572-9611]{Silva J\"arvinen}
\affiliation{Leibniz-Institut f\"ur Astrophysik Potsdam (AIP), An der Sternwarte~16, 14482~Potsdam, Germany}

\author[0000-0002-1015-3268]{Donald W. Kurtz}
\affiliation{Centre for Space Research, Physics Department, North West University, Mahikeng 2735, South Africa}
\affiliation{Jeremiah Horrocks Institute, University of Central Lancashire, Preston PR1 2HE, UK}

\author[0000-0003-2058-6662]{George R. Ricker}
\affiliation{MIT Kavli Institute and Department of Physics, 77 Massachusetts Avenue, Cambridge, MA 02139} 

\begin{abstract}
We report the detection and characterization of a new magnetospheric star, HD\,135348, based on photometric and spectropolarimetric observations. The \TESS~light curve of this star exhibited variations consistent with stars known to possess rigidly rotating magnetospheres (RRMs), so we obtained spectropolarimetric observations using the Robert Stobie Spectrograph (RSS) on the Southern African Large Telescope (SALT) at four different rotational phases. From these observations, we calculated the longitudinal magnetic field of the star $\langle B_z \rangle$, as well as the Alfv\'en and Kepler radii, and deduced that this star contains a centrifugal magnetosphere. However, an archival spectrum does not exhibit the characteristic ``double-horned'' emission profile for H$\alpha$ and the Brackett series that has been observed in many other RRM stars. This could be due to the insufficient rotational phase coverage of the available set of observations, as the spectra of these stars significantly vary with the star's rotation. Our analysis underscores the use of \TESS~in photometrically identifying magnetic star candidates for spectropolarimetric follow-up using ground-based instruments. We are evaluating the implementation of a machine learning classifier to search for more examples of RRM stars in \TESS~data.
\end{abstract}


\section{Introduction}
\label{sec:intro}
The magnetic fields generated by hot, massive stars (types O, B, and A) have significant downstream effects. They can affect the surface rotation rates via magnetic braking \citep{1967ApJ...148..217W, 2008MNRAS.385...97U}, introduce chemical abundance inhomogeneities and peculiarities \citep{1999A&A...351..554H}, and confine the stellar wind in a magnetosphere \citep{1984ApJ...282..591F, 1997A&A...323..121B, 2002ApJ...576..413U}. In such stars, the wind is driven along the field lines toward the magnetic equator -- leading to a strong shock when the wind components of the two hemispheres collide. This heats the plasma to X-ray temperatures, a phenomenon that can be described by the magnetically confined wind shock model of \citet{1997A&A...323..121B}.

Figure 3 of \citet{2013MNRAS.429..398P} presents a classification scheme (the ``magnetic confinement v/s rotation diagram'', henceforth MCRD) for stars that possess magnetospheres. This can be used to differentiate between two types of magnetospheres -- centrifugal and dynamical; these classifications are based on two characteristic radii: the Alfv\'en radius, $R_A$, and the Keplerian co-rotation radius, $R_K$.\footnote{These are often also respectively referred to as $R_m$, for the magnetospheric radius, and $R_c$, for the corotation radius.} Beyond $R_A$, the radial stellar wind forces the magnetic field lines to also become radial (see \citealt{1969SoPh....9..131A}, and references therein).

A star has a dynamical magnetosphere when $R_K>R_A$; here, material confined in closed magnetic loops falls back onto the star's surface on a dynamical timescale due to gravity \citep{2012MNRAS.423L..21S}. On the other hand, a star has a centrifugal magnetosphere (CM) when $R_A > R_K$, i.e., material caught between $R_A$ and $R_K$ is supported against radial infall by a centrifugal force. This causes the plasma to build up, creating a relatively dense magnetosphere \citep{2013MNRAS.429..398P}.

In some early B stars, the combination of rapid rotation and a strong magnetic field leads to the formation of a centrifugally supported magnetosphere with rotationally modulated hydrogen line emission. These stars, which can be accounted for by the rigidly rotating magnetosphere (RRM) model \citep{2005MNRAS.357..251T}, are rare. The first one detected, $\sigma$~Ori E, was identified due to emission variability in hydrogen lines (specifically, H$\alpha$ and the Brackett series, i.e., transitions to the $n=4$ level of the H atom). The structure of the spectral features suggests that gas is trapped in magnetospheric clouds around the star, a conclusion derived from extensive modeling of its observational signatures by \citet{2005MNRAS.357..251T}. Ten such stars have been identified as exhibiting a ``double-horned'' emission pattern at the H$\alpha$ wavelength (see the Introduction of \citealt{2020MNRAS.499.5379S}, and references therein). These RRM stars are rather mysterious; their rotational velocities can approach breakup velocity, but their spin-down timescales via magnetic braking are significantly shorter than their estimated ages \citep{2014ApJ...784L..30E}.

The recent proliferation of large-scale, continuously observing space-based surveys has significantly increased the chance of identifying stars with detectable flux modulations potentially arising from a magnetosphere. Photometric missions such as the \Kepler~space telescope and the Transiting Exoplanet Survey Satellite (\TESS\,) have discovered several new classes of stars via their photometric modulation signatures; these techniques are easily extensible to the light curves of RRM stars (for models of these, see, e.g., \citealt{2016MNRAS.462.3830O, 2016A&A...594A..75K}). Thus, large-scale searches are underway that aim to identify and characterize rotational modulation profiles of stars observed by \Kepler~(see, e.g., \citealt{2013ASSP...31..247B,2016MNRAS.463.1740B}) and \TESS~(see, e.g., \citealt{2019MNRAS.487..304D, 2019MNRAS.487.4695S}).

In this Letter, we present HD\,135348, with spectral type B3V (from \citealt{1969ApJ...157..313H}), as a new magnetospheric B star identified through photometric observations obtained by \TESS. We present spectropolarimetric observations of this star, along with a determination of its longitudinal magnetic field $\langle B_z \rangle$. Finally, we locate HD\,135348 on the MCRD, and discuss this star in the context of other B stars with magnetospheres.
\section{Observations}
\label{sect:obs}
\subsection{TESS Observations}
\begin{figure}
    \begin{center}
    \includegraphics[width=\linewidth]{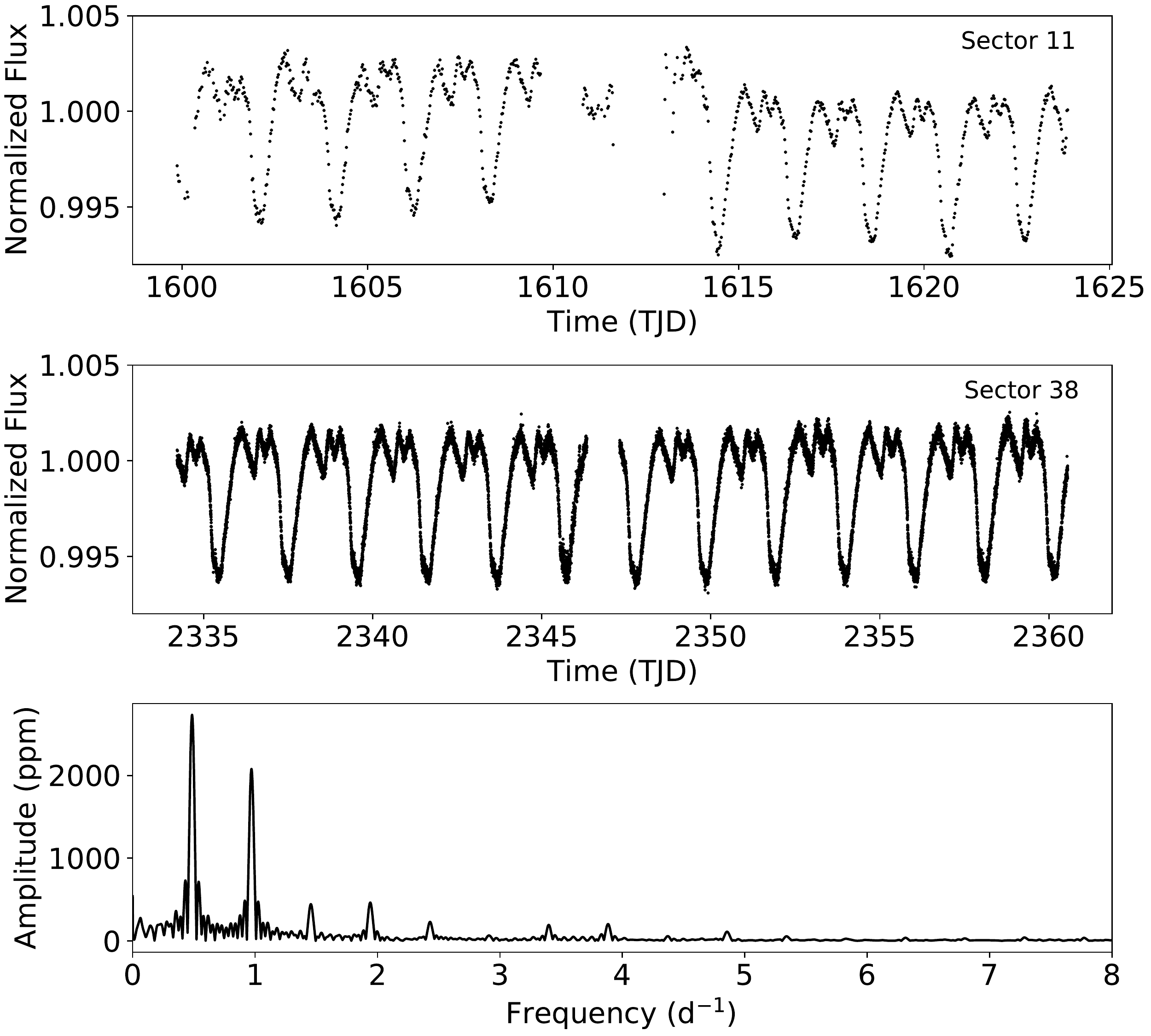}
    \caption{The top two panels show \TESS~light curves for HD\,135348 from Sectors 11 and 38. The bottom panel contains a Discrete Fourier Transform of the Sector 38 data, showing a period of 2.0593\,d. The data gap in the middle of the light curves is caused by the lack of observations during data downlink. Approximately 5\% of the data points in the Sector 11 light curve were clipped. The removed points arise from scattered light entering the aperture; the deviations of these points from unity were found to be much greater than the semi-amplitude of the periodic flux variations ($\sim$5 ppt).}
    \label{fig:tess_lc_hd135}
    \end{center}
\end{figure}

HD\,135348 (TIC\,142505974) was observed by \TESS~during Sectors 11 and 38, which lasted from 2019 April 22 to 2019 May 21, and 2021 April 28 to 2021 May 26, respectively. During Sector 11, it was observed at 30-min cadence in the full-frame images (FFIs), while it was observed at 2-min cadence in Sector 38. The Sector 38 light curve is available in both SAP (simple aperture photometry) and PDCSAP (presearch data conditioning SAP) forms, which were generated by processing via the SPOC pipeline at the NASA Ames Research Center \citep{jenkinsSPOC2016}. We used the PDCSAP data for our analysis. We also used the \texttt{eleanor} package \citep{2019ascl.soft04022F} to extract a light curve from the Sector 11 FFIs. The two light curves are shown in Figure\,\ref{fig:tess_lc_hd135}.

Visual inspection indicates that the light curves in Figure\,\ref{fig:tess_lc_hd135} exhibit significant periodic variability. Taking a Discrete Fourier transform (see, e.g., \citealt{1985MNRAS.213..773K}), the results of which are presented in the bottom panel of Figure \ref{fig:tess_lc_hd135}, we find a period of 2.0593~$\pm$~0.0001\,d. The additional peaks visible in the Fourier transform are harmonics of this fundamental frequency.

\begin{figure}
    \centering
    \includegraphics[width=\linewidth]{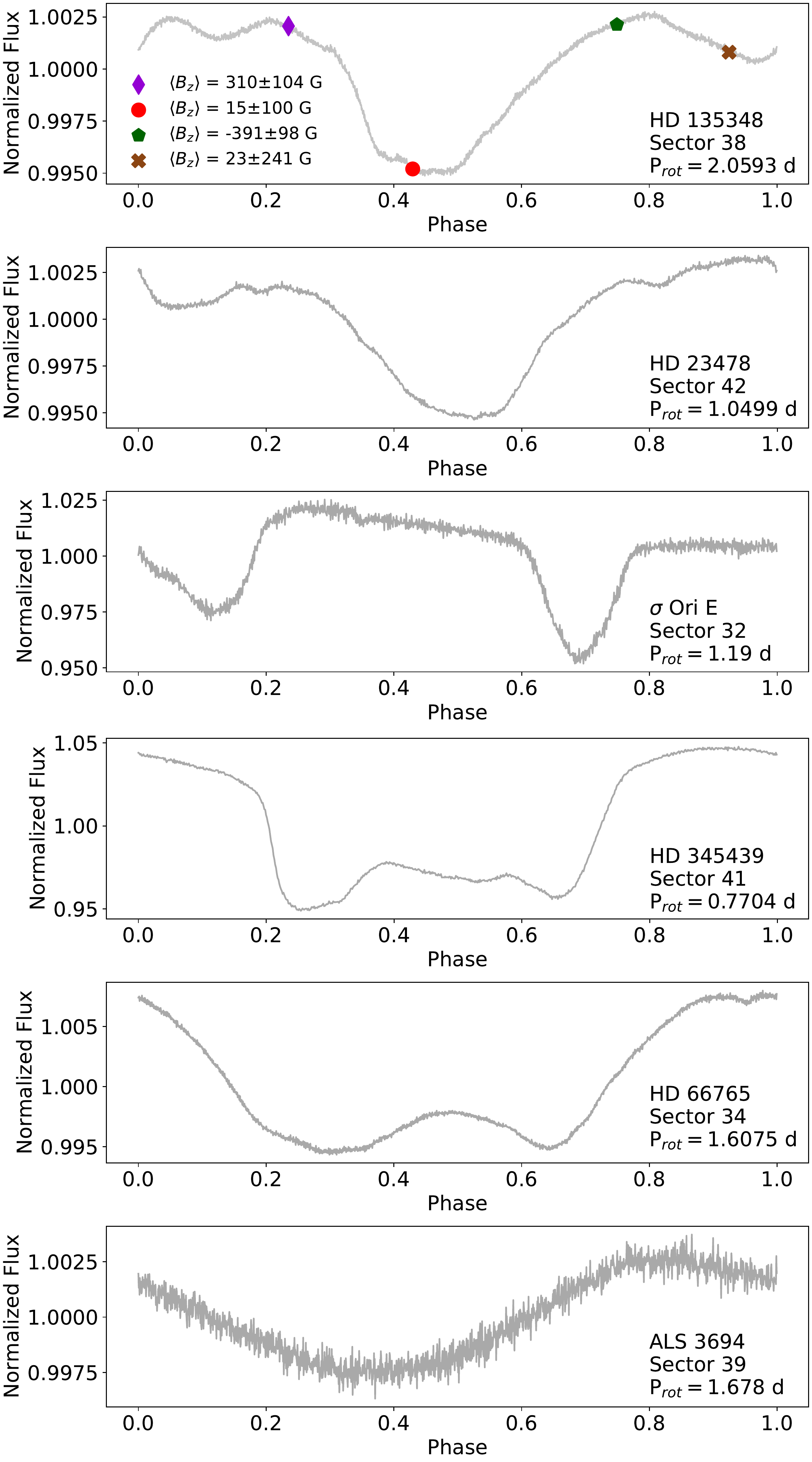}
    \caption{A comparison of binned, phase-folded \TESS~light curves for various magnetic stars with that of HD\,135348 in the top panel. Its light curve most closely resembles that of the known RRM star HD\,23478, shown in the second panel. The light curves of two other confirmed RRM stars, $\sigma$~Ori E and HD\,345439, are shown in the third and fourth panels. The fifth panel shows the light curve of HD\,66765, a star that shows H$\alpha$ in emission (a strong indicator that an RRM exists), and the bottom panel shows the light curve of a CM star, ALS 3694. The markers on the plot for HD\,135348 represent the rotational phases at which the magnetic field measurements were made (see Section~\ref{subsec:salt} and Table~\ref{tab:meas}). We emphasize the significant differences in the light curve morphologies for CM and RRM stars. These light curves are binned to 0.002~units of phase. Note that the y-axis scales vary in each panel, in order to more clearly show the individual stars' flux variations. The scatter visible in the light curve of ALS\,3694 is likely not astrophysical in origin, and may simply be rms scatter.}
    \label{fig:tess_rrm_lcs}
\end{figure}

The light curve of HD\,135348 looks similar to other magnetospheric stars observed by \TESS~, especially HD\,23478, which is presented in the second panel of Figure\,\ref{fig:tess_rrm_lcs}. The third and fourth panels of Figure\,\ref{fig:tess_rrm_lcs} show two other well-known RRM stars and highlight the diverse morphologies of the light curves of these stars. The fifth panel shows the light curve of HD\,66765, the star located closest to HD\,135348 on the MCRD (discussed further in Section \ref{sec:magnetosphere_calc}); the bottom panel contains the light curve of a CM star, ALS\,3694. Comparing the top 5 panels of this plot to the bottom one emphasizes the distinct morphologies of RRM and CM stars' light curves, especially the large, noticeable dip(s) in flux exhibited by RRM stars; these features could enable us to photometrically identify such stars. This figure also contains the first published \TESS~light curves of RRM stars.

While the rotational period of HD\,135348 is above average for RRM stars (e.g., HD\,23478 has $P_{\rm rot}$ = 1.05\,d; $\sigma$\,Ori\,E has $P_{\rm rot}$ = 1.19\,d), it is not out of the realm of possibility: CPD$-62^{\circ}2124$, mentioned in \citet{2020MNRAS.499.5379S} as having H$\alpha$ emission consistent with the RRM model, has $P_{\rm rot}$ = 2.628~d \citep{2017MNRAS.472..400H}.

\subsection{SALT RSS Spectropolarimetry}
\label{subsec:salt}
The determination of magnetic fields in early B-type stars is usually based on measurements of the mean longitudinal magnetic field, i.e., of the line-of-sight field component averaged over the visible stellar hemisphere, using circularly polarized light. To search for a magnetic field in HD\,135348, we obtained four low-resolution spectropolarimetric observations with the Robert Stobie Spectrograph (RSS; \citealt{2003SPIE.4841.1463B}), mounted on the Southern African Large Telescope (SALT; \citealt{2006SPIE.6267E..0ZB}). For more details about the RSS and how its data could be processed, we direct the reader to \citet{2010SPIE.7735E..17B}, who discuss the first attempts at spectropolarimetric data reduction with SALT, and \citet{2016SPIE.9908E..2KP}, who discuss recent updates to the detector.

To probe the wavelength range of interest (4424\,\AA~$< \lambda <$~5090\,\AA), we used grating PG3000 at an angle of 46.625$^\circ$ and a camera angle of 93.25$^\circ$. With a 0.6\,arcsec slitwidth, we achieved a spectral resolution of $R\sim 9600$. At each epoch, we took a continuous series of eight exposures using the standard readout mode. The quarter waveplate was oriented at a 45$^{\circ}$ angle for exposures 1, 4, 5, and 8, and at 135$^{\circ}$ for exposures 2, 3, 6, and 7. Each observation consisted of a 320\,s integration (8~$\times$~40\,s) in order to achieve a S/N~$\gtrsim$~1000 per exposure.

We obtained observations of HD\,135348 at four different epochs to sample different rotation phases (the zero point for the phases was chosen to coincide with the midpoint of the \TESS~observation, at BJD\,2459347.88390; see Table~\ref{tab:meas} for the list of phases). This aimed to maximize the field detection probability and avoid missing the magnetic field due to a potentially unfavorable viewing angle at specific phases. The chemically peculiar A9VpSrCrEu star $\gamma$\,Equ (HD\,201601), with an extremely long rotation period of $>95.5$\,yr \citep{2016MNRAS.455.2567B, 2018AstBu..73..463S} and a well-characterized magnetic field, was used as a magnetic standard star. With $m_{\rm V}=4.7$, this bright star is frequently used to test and characterize instrumental polarimetric capabilities in both the Northern and Southern Hemispheres.

The first steps in the data reduction are standard processes for spectroscopic data; all raw frames were reduced using {\sc{starlink}} software \citep{2014ASPC..485..391C} to apply bias correction, cosmic ray masking, and flat fielding. The wavelength calibration was then carried out using CuAr arc lamp exposures obtained after the science frames. The wavelength solution for the O beam was applied to both the O and E beams in all observed spectra to ensure no drift in the wavelength solution.

The method to assess the presence of a magnetic field is very similar to that done using the European Southern Observatory (ESO) instruments FORS\,1 and 2 in their spectropolarimetric mode. These techniques have been presented in many prior publications (see, e.g., \citealt{2004A&A...415..661H, 2004A&A...415..685H}, and references therein). We note that other works have made use of the {\sc{polsalt}} reduction pipeline\footnote{\url{https://github.com/saltastro/polsalt}}; however, this code cannot yet handle circular polarization data. The longitudinal magnetic field measurements are listed in Table~\ref{tab:meas}.

\begin{table}
\caption{Longitudinal magnetic field values $\langle B_z \rangle$ obtained for $\gamma$\,Equ and HD\,135348 from RSS observations. The first column shows time of mid-exposure, followed by the signal-to-noise ratio ($S/N$) of the RSS Stokes~$I$ spectra measured near 4480\,\AA. The measurements of $\langle B_z \rangle$ using the Monte Carlo bootstrapping test and the null spectra are presented in Columns~3 and 4. Finally, in Column~5, we indicate the rotational phase at which the measurement was taken. The zero point for the phases is BJD\,2459347.88390, corresponding to the approximate midpoint of the \TESS~observation.}
\label{tab:meas}
\begin{center}
\begin{tabular}{ccr@{$\pm$}lr@{$\pm$}lc}
\hline
\multicolumn{1}{c}{Date} & 
\multicolumn{1}{c}{$S/N$} & 
\multicolumn{2}{c}{$\left<B_{\rm z}\right>$} &
\multicolumn{2}{c}{$\left<B_{\rm z}\right>_N$} &
\multicolumn{1}{c}{Phase} \\
\multicolumn{1}{c}{\textit{(BJD-2400000)}} &
\multicolumn{1}{c}{} &
\multicolumn{2}{c}{\textit{(G)}} &
\multicolumn{2}{c}{\textit{(G)}} &
\multicolumn{1}{c}{} \\
\hline
\multicolumn{7}{c}{$\gamma$\,Equ} \\
\hline
59437.43915 & 1349 &$-$638 & 36 &$-$106& 34 & N/A \\
\hline
\hline 
\multicolumn{7}{c}{HD\,135348} \\
\hline
59437.31835 & 1978 & 15 & 100 &$-$25 & 109 & 0.430 \\
59438.33863 & 1047 & 23 & 241 & 8 & 238 & 0.925 \\
59448.27340 & 1740 &$-$391 & 98 &$-$20 & 105 & 0.749 \\
59449.27366 & 1573 & 310 & 104 & 41 & 111 & 0.235 \\
\hline
\hline
\end{tabular}
\end{center}
\end{table}
The mean longitudinal magnetic field $\langle B_{\rm z} \rangle=-638\pm36$\,G found for $\gamma$\,Equ is consistent with recent measurements presented in Figure 1 of \citet{2021MNRAS.tmpL..86H}, who observed this star using the high-resolution Potsdam Echelle Polarimetric and Spectroscopic Instrument (PEPSI), located at the Large Binocular Telescope. Specifically, this agrees with their measurement from 2019 May -- $\langle B_{\rm z} \rangle = -663\pm21$\,G -- and is generally consistent with the overall predicted variability of this star's mean magnetic field $\langle B \rangle$ throughout time.

The strongest mean longitudinal magnetic field for HD\,135348, $\langle B_{\rm z}\rangle = -391\pm98$\,G, is measured at a significance level of 4.0$\sigma$ at the rotation phase 0.749. After about half a rotation cycle, at phase 0.235, we measure a positive longitudinal field $\langle B_{\rm z} \rangle = 310\pm104$\,G, at a significance level of 3.0$\sigma$. The detection of a magnetic field in HD\,135348 makes this star a good candidate for long-term spectropolarimetric monitoring to fill in the gaps of its full magnetic phase curve. 

\section{Spectral variability}
\label{sec:var}
\begin{figure*}
    \begin{center}
    \includegraphics[width=\linewidth]{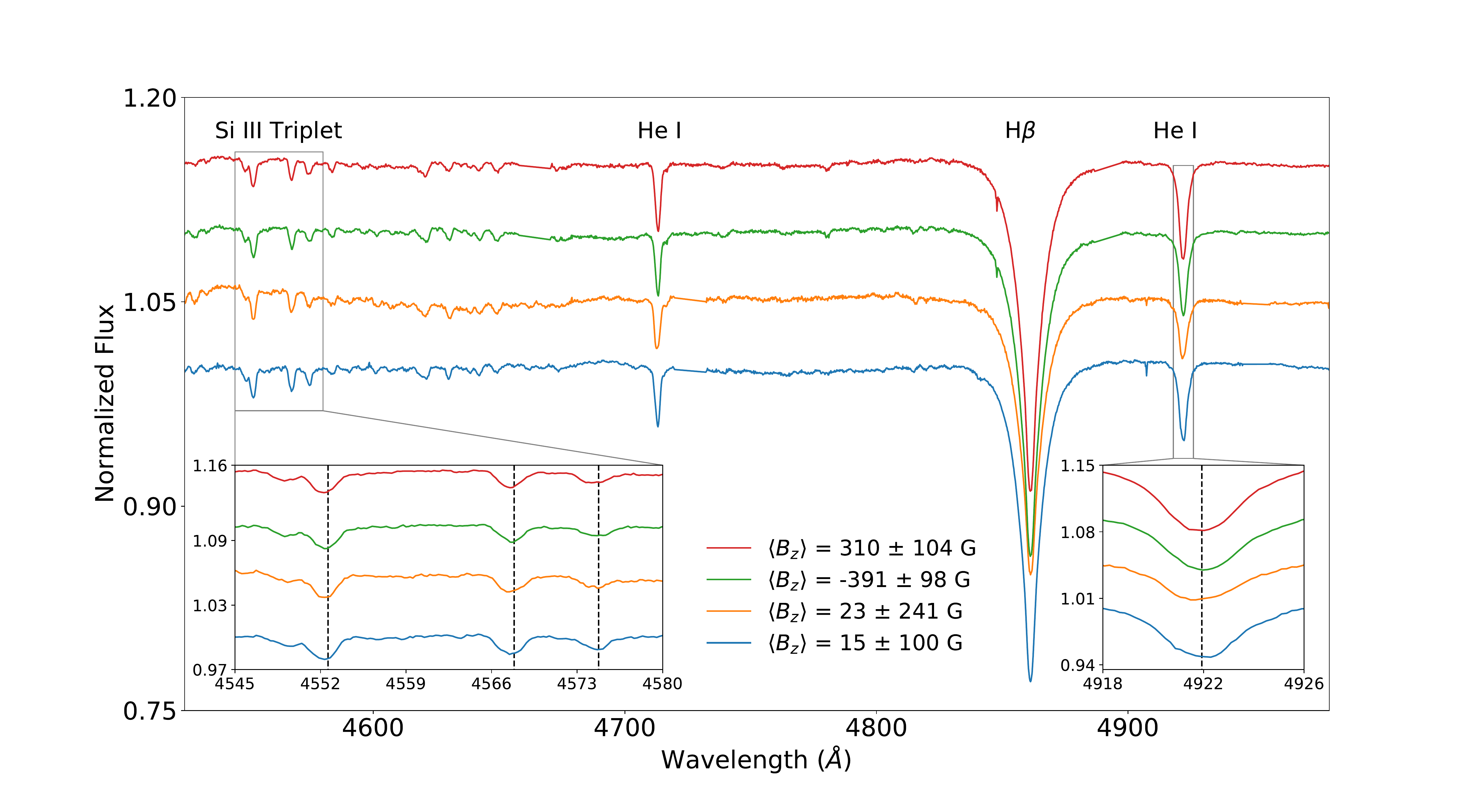}
    \caption{A plot of four Stokes~$I$ spectra taken with the RSS over four different epochs with insets showing the variability of the line profiles of the Si\,{\sc{iii}} triplet at 4552.62\,\AA, 4567.84\,\AA, and 4574.76\,\AA, along with the profiles of He\,{\sc{i}} at 4921.93\,\AA. These wavelengths have been labeled with dotted vertical lines in both inset plots. Three of the four spectra have been offset for better visibility. We have labeled two other key spectral lines (another He {\sc{i}} line, and also H$\beta$)  visible in the plot.}
    \label{fig:spectra}
    \end{center}
\end{figure*}

RSS Stokes~$I$ spectra showing the variability of different spectral lines are presented in Figure\,\ref{fig:spectra}. To normalize each spectrum, we fit a fifth-order polynomial to the overall spectrum (masking out the largest and most relevant spectral lines), and then subtracted this fit from the raw data. The asymmetries in certain spectral line profiles, such as those of He\,{\sc{i}}, arise from the inhomogeneous surface distribution of the corresponding elements, giving rise to chemical spots. As a result, the absorption varies with stellar rotation and manifests as a pronounced asymmetry in the corresponding spectral line (see, e.g., the bottom right inset plot in Figure\,\ref{fig:spectra}, showing a large asymmetry in the He\,{\sc{i}} line at 4922\,\AA). 

While we attempted to calculate the radial velocity (RV) of the star, our accuracy was limited by the low resolution of the spectra and insufficient phase coverage. However, as shown in the inset in the bottom right of Figure~\ref{fig:spectra}, we observe significant variations in the He\,{\sc{i}} line profile at the four observed phases, along with shifts in the RV. This behavior is expected for magnetic B type stars, especially those with surface He and Si spots. Archival RV values, from \citet{1953GCRV..C......0W} and \citet{2006AstL...32..759G}, indicate an average value of $-21 \pm 3.5$ km s$^{-1}$. A crude calculation based on our observations suggests a mean RV of $\sim -$~30\,km\,s$^{-1}$ (with a large uncertainty). This value is in line with prior measurements; we plan to take further observations for a more precise estimate.

Finally, we used two different techniques to characterize the star's rotational velocity and determine $v\,\sin\,i$. First, we used the mean profiles of the Si\,{\sc{iii}} triplet lines observed at the four different epochs. This involved fitting Gaussians to these mean profiles to calculate their FWHM. This method yields an estimate for $v\,\sin\,i$ of 125 $\pm$ 9\,km\,s$^{-1}$. To verify this value, we selected a sample of stars from \citet{shultz2018} with similar spectral type (B2-B5) and magnetic field strength. We downloaded high-resolution spectra ($R\approx65,000$) from the Echelle SpectroPolarimetric Device for the Observation of Stars at the Canada-France-Hawaii Telescope \citep{2006ASPC..358..362D}. These archival spectra were degraded to the resolution of the RSS by convolving them with a Gaussian profile. This enabled us to estimate that $v\,\sin\,i$ is $120\pm10$~km\,s$^{-1}$. Our $v\,\sin\,i$ estimates are on the higher end for magnetic B stars, when compared to those in Table~1 of \citet{2013MNRAS.429..398P}. Consequently, we can conclude that this star is rapidly rotating, leading to a significantly smaller Keplerian co-rotation radius.

\section{Estimating Stellar Parameters}
\label{sec:est-stel-param}
The \TESS~Input Catalog \citep{2019AJ....158..138S} does not have reliable stellar parameters for HD\,135348, and the online SIMBAD catalog \citep{Wenger2000} only contains information about the parallax, radial velocities, and inferred spectral type. As a result, we sought to derive a reliable estimate for these values to use for our calculations of magnetospheric properties, in Section \ref{sec:magnetosphere_calc}.

This star is bright, with $m_{\rm V} = 6.05$ and $T_{\rm mag} = 6.184$\footnote{$T_{\rm mag}$ refers to the magnitude in the \TESS~band (6000-10000\,\AA).} \citep{2019AJ....158..138S}. Typically, we would use an InfraRed Flux Method (IRFM) to calculate the temperature from a measured color index (see, e.g., \citealt{2020RNAAS...4...52M}). However, the presence of chemical spots on B stars leads to significant flux redistribution from the far-UV region to the near-UV and visible regions of the spectrum (see, e.g., Section 5 in \citealt{2013A&A...556A..18K}). This therefore precludes us from using broad-band color measurements, such as the $G-RP$ and $BP-G$ indices from Gaia. Moreover, many of the published IRFM coefficients are not valid for stars with $T_{\rm eff} \gtrsim 8000$~K. As a result, we use archival uvby$\beta$ colors (which are narrow-band) and estimate $T_{\rm eff}$ by interpolating within the grid of \citet{1985MNRAS.217..305M}\footnote{\url{https://wwwuser.oats.inaf.it/castelli/colors/uvbybeta.html}}.

The most recent uvby$\beta$ measurements are from \citet{2015A&A...580A..23P}. Two parameters, $\beta$ and $c_1$, are enough to estimate $T_{\rm eff}$; for HD\,135348, $\beta = 2.661 \pm 0.006$, and $c_1 = 0.326 \pm 0.01$. These measurements are consistent with archival data from \citet{1977A&AS...27..443G} and \citet{1998A&AS..129..431H}, but we add the caveat that such narrow-band color indices, while far more reliable, are still not immune from the photometric effects of chemical spots in B stars explored in \citet{2013A&A...556A..18K} -- emphasizing the importance of repeated, long-term measurements for these stars to obtain full phase coverage.

An estimate derived from the grid of \citet{1985MNRAS.217..305M} yields $T_{\rm eff} \approx 16\,000 \pm 1\,000$~K, assuming solar metallicity and a microturbulent velocity $\xi$ of 2.0 km\,s$^{-1}$ (many early- to intermediate-type B stars have $\xi \sim 1-10$~km\,s$^{-1}$ -- see, e.g., \citealt{2010A&A...515A..74L}). We use the Gaia parallax of $3.17\pm0.07$~mas\footnote{We note that there are significant, well-documented issues with Gaia parallaxes of stars with $m_V \lesssim 5$ (see, e.g., \citealt{Drimmel_2019}); however, HD\,135348 is far enough below this cutoff that we can take the parallax estimate to be reliable.} \citep{2021A&A...649A...2L} and this star's $m_V$ to calculate an absolute magnitude $M_V = -1.44\pm0.05$. However, to convert this to a luminosity, we have to apply an appropriate bolometric correction as detailed in \citet{1998A&A...333..231B}, assuming $log\,g = 4.0$. This yields $M_{\rm V,bol} = -2.84\pm0.05$, and a luminosity $L = 1076\pm50$~L$_\odot$. Then, we apply the Stefan-Boltzmann law to derive $R = 4.27 \pm 0.01$~R$_\odot$. Finally, we used the mass-luminosity relationship for intermediate-mass stars presented in Table 6 of \citet{2007MNRAS.382.1073M} to obtain $M_* = 5.66_{-0.08}^{+0.06}$\,M$_\odot$. This yields $\log\,g = 3.93 \pm 0.01$.

Using our derived parameters and the tables of \citet{2013ApJS..208....9P}, we identify this star's spectral type as B2.5-4V. This agrees with the spectral type of B3V in \citet{1969ApJ...157..313H}. We additionally emphasize that interstellar reddening does not significantly affect our color estimates, as the extinction coefficient $A_G = 0.06$. Our inferred spectral type agrees with those of other known RRM stars, which are all early- to intermediate-B stars.

\section{The Magnetosphere of HD\,135348}
\label{sec:magnetosphere_calc}
\subsection{Is the Magnetosphere Centrifugal or Dynamical?}
First, we estimate $R_A$ and $R_K$. The former is derived from the magnetic wind confinement parameter $\eta_*$:
\begin{equation}
    \eta_* = \frac{B_{\rm eq}^2 R_*^2}{\dot{M}_{B=0} V_{\infty}}.
\end{equation}
We can use the definitions and equations from \citet{2013MNRAS.429..398P} to calculate $\eta_*$. Firstly, $B_{\rm eq} \equiv B_d/2$, where $B_d$ is thrice the largest longitudinal magnetic field measurement (in our case, $\sim$1.2\,kG). Also, $\dot{M}_{B=0}$ and $V_{\infty}$ are the fiducial mass-loss rate and the ``terminal speed'' that the stellar wind would have with no magnetic field:
\begin{equation}
    \dot{M}_{B=0} =0.551 \frac{L_*}{c^2} \left(\frac{1000\Gamma_e}{1-\Gamma_e}\right)^{0.818}
\end{equation}
\begin{equation}
    V_{\rm esc} \equiv \left(\frac{2GM_* \left(1 - \Gamma_e\right)}{R_*}\right)^{1/2}.
\end{equation}
Here, $\Gamma_e = \kappa_e L_* / 4\pi G M_* c$ is the Eddington parameter for the electron scattering opacity, $\kappa_e$. We thus obtain $\eta_* = 8.78 \pm 2.99 \times 10^3$. We substitute $\eta_*$ to find $R_A$:
\begin{equation}
    \frac{R_A}{R_*} \approx 0.3 + \left(\eta_* + 0.25\right)^{1/4}.
\end{equation}
This yields $R_A \approx 9.98\pm 0.82$\,R$_*$.

We find $R_K$, given by
\begin{equation}
    R_K \equiv \left( \frac{GM}{\omega^2}\right)^{\frac{1}{3}},
\end{equation}
to be $10.86 \pm 0.02$\,R$_\odot$. This implies that $R_K/R_* \sim 2.84$.

With estimates for $R_A$ and $R_K$, we can see that $R_A>R_K$ and thus place HD\,135348 on the MCRD in the region containing stars possessing a centrifugal magnetosphere. Its closest neighbor on the MCRD is HD\,66765, which was found to exhibit (relatively weak) H$\alpha$ emission lines (see Figure B1 in \citealt{2020MNRAS.499.5379S}). However, the fact that this star has both H$\alpha$ emission and a light curve reminiscent of other RRM stars (see Figure \ref{fig:tess_rrm_lcs}) strongly suggests that this star also possesses an RRM. Additionally, \citet{2020MNRAS.499.5379S} claim that the RRM model can adequately explain the observed H$\alpha$ emission of stars in this region of the MCRD, implying that such a model can explain our photometric observations of HD\,135348.

Indeed, all that is needed for an RRM to exist is that $R_A > R_K$, which we have proven for HD\,135348. \citet{2014ApJ...784L..30E} state that RRMs are more likely to occur in stars with magnetic fields $>10$\,kG, but are not improbable in stars with dipole magnetic fields of $\sim1$\,kG -- such as HD\,135348. We thus sought to directly compare key spectral features of RRM stars to archival spectra of HD\,135348 to evaluate this claim.

\subsection{Comparison to Known RRM Stars}
We used archival observations from the XSHOOTER and UVES spectrographs \citep{2011A&A...536A.105V, 2000SPIE.4008..534D} on the Very Large Telescope (VLT). These spectrographs enable us to investigate two regions that we could not study with our chosen RSS configuration: H$\alpha$ ($\lambda=6562.3$~\AA), and the Brackett series, for $n \gtrsim 10$. The XSHOOTER spectrum for HD\,135348 ($R \sim 5040$, S/N~$\sim 260$) was obtained on 2013 March 2, with an exposure time of 5\,s, and spans 5340 to 21000\,\AA. We compared this to observations of HR\,5907, a known RRM star, that were taken using UVES on 2010 April 13 and XSHOOTER on 2014 April 29. For all spectra, we used the processed version from the ESO Science Portal.

\begin{figure}
    \centering
    \includegraphics[width=\linewidth]{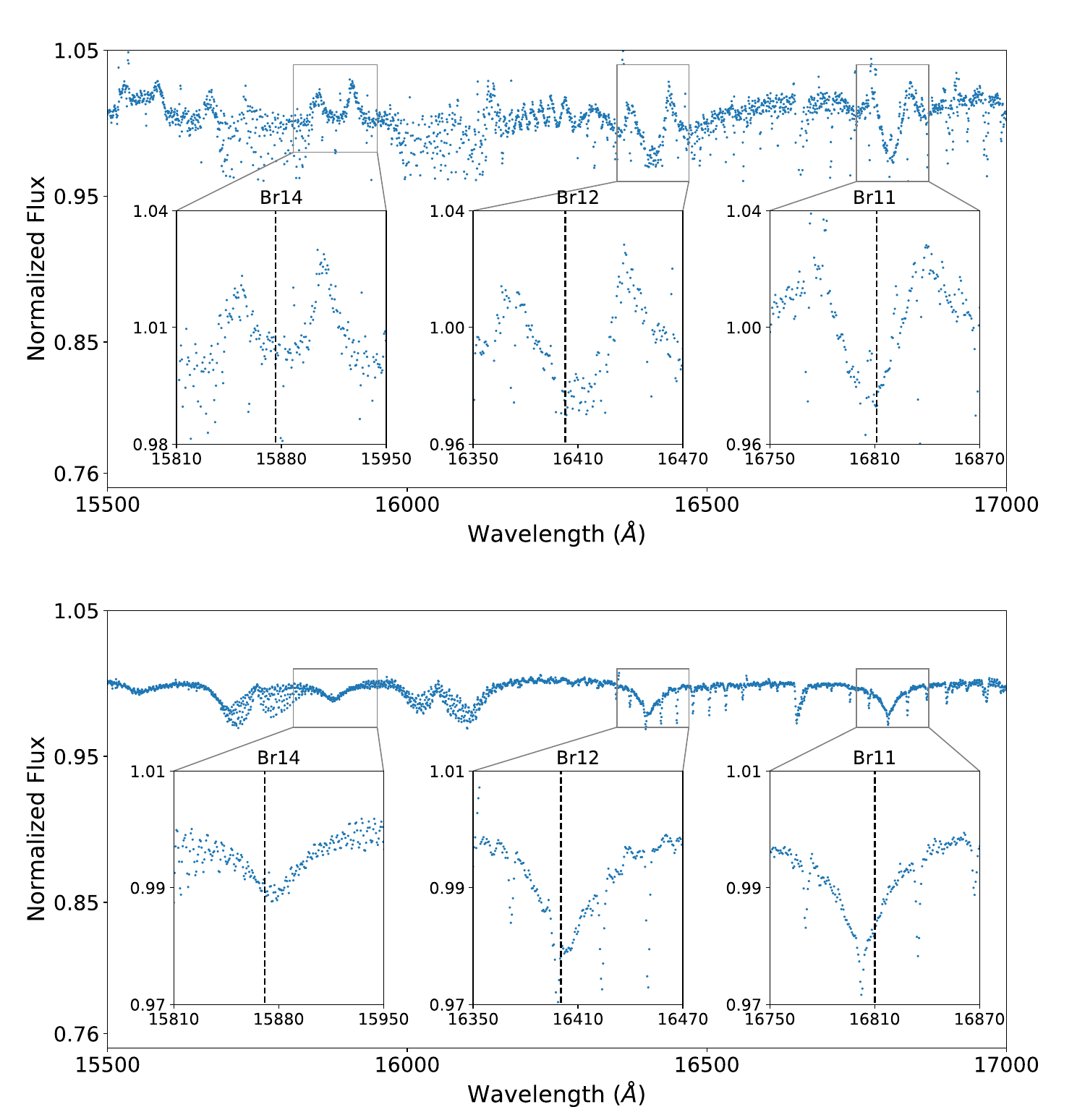}
    \caption{Archival XSHOOTER spectra showing ``double-horned'' Brackett line emission for HR\,5907 (top panel), compared with the lack of a similar profile in HD\,135348 (bottom panel). A similar plot of the Brackett series, albeit for HD\,23478 and HD\,345439, can be found in figure\,1 of \citet{2014ApJ...784L..30E}. The dotted vertical lines mark the wavelengths of the corresponding Brackett series line (Br14: 15876.3 \AA, Br12: 16402.8 \AA, Br11: 16802 \AA).}
    \label{fig:brackett}
\end{figure}

In contrast to typical RRM stars, the archival spectrum of HD\,135348 does not show an emission profile at H$\alpha$, or throughout the Brackett series; this, however, does not exclude the possibility of emission at different rotation phases. Known RRM stars, such as HR\,5907, exhibit a characteristic ``double-horned'' emission-line profile throughout the Brackett series, especially for $n > 10$ (see Fig. 1 in \citealt{2014ApJ...784L..30E}). In Figure\,\ref{fig:brackett}, we show a comparison of the Brackett series for HR\,5907 and HD\,135348; the archival spectrum shows only absorption features for the latter star. 

Some other RRM stars exhibit a similar double-horned symmetric profile at the core of the H$\alpha$ line, but this is not universal across known RRM stars (e.g., there is a pronounced asymmetry in the ``horns'' of the line profile in the spectrum of HD\,345439 -- see the inset in Fig. 3 of \citealt{2014ApJ...784L..30E}). In a number of RRM stars, the H$\alpha$ emission is also highly variable and could disappear at some rotational phases. Visual inspection of the H$\alpha$ region of the spectra for both HR\,5907 and HD\,135348 revealed a ``double-horned'' profile for HR\,5907, but only an absorption feature in the spectrum of HD\,135348. As a result, we claim that this is a CM star, but cannot definitively conclude whether this is an RRM star, due to  the insufficient phase coverage of available spectroscopic observations.

\citet{2020MNRAS.499.5379S} postulate that RRMs may be a subclass of CM stars, albeit with with a favorable viewing angle. The stellar wind plasma accumulates in a torus around the star's magnetic equatorial plane. When the clouds of plasma transit the star, we observe H$\alpha$ absorption; in stars where the centrifugal magnetosphere is seen face-on, the emission of an RRM peaks at $R_K$, which leads to the characteristic double-horned profile. On the other hand, \citet{2020MNRAS.499.5366O} suggest that a lack of emission around H$\alpha$, which is commonly observed in stars with spectral type later than $\sim$B6 ($T_{\rm eff} \lesssim 16$~kK and L~$\lesssim 800$~L$_{\odot}$), may arise from leakage that prevents filling of the CM to the level needed for H$\alpha$ emission (because of a lower $\dot{M}$) or the predominance of a metal-ion driven wind, which would lack the hydrogen needed for emission. HD\,135348, which we infer to be an intermediate B star in Section~\ref{sec:est-stel-param}, could possess such an underdense centrifugal magnetosphere -- a potential explanation for the lack of observed hydrogen emission.

The major implication of our study is the fact that we can potentially photometrically identify RRM candidates from large-scale sky surveys. With the vast amounts of data we will obtain as large sky surveys (such as the Vera Rubin Observatory) come online, and space-based surveys such as \TESS~continue observing large parts of the sky, we could train a machine learning algorithm to search for characteristic features of RRM light curves using the model described in \citet{2008MNRAS.389..559T}. 

\section{Conclusions}
\label{sec:conc}
In this Letter, we use \TESS~data and spectropolarimetric observations to characterize a magnetic B star, HD\,135348. Our observations and calculations allow us to deduce that this star possesses a centrifugal magnetosphere and could potentially be an RRM star. However, we cannot make a conclusive determination about the latter point due to the small number of magnetic field measurements, which neither provide full rotational phase coverage nor enable us to deduce the field strength and its geometry. Future work will entail obtaining additional spectropolarimetric observations to construct this star's full magnetic phase curve, enabling us to either verify or discount the presence of an RRM.

We note that this is the first publication to present \TESS~photometry of RRM stars. \TESS's short-cadence continuous observations have enabled us to precisely constrain stellar rotation rates. Further \TESS~data on any RRM star can also constrain any variability in the light curve between cycles, as these stars have comparatively short rotational periods ($\lesssim 2.5$\,d), and a single \TESS~observing sector covers several rotation cycles. We are also developing a machine learning classifier to identify high-likelihood RRM star candidates in \TESS~data for spectropolarimetric follow-up.
\section*{Acknowledgements}
Funding for the \TESS~Mission comes from the NASA Science Mission Directorate. RJ would like to acknowledge funding from the MIT Department of Physics as a Graduate Research Assistant. This work was based on observations made with SALT under program ID 2021-1-DDT-002 (PI: Holdsworth). Archival observations obtained from the ESO were conducted under program ID 093.D-0448 (PI: Shultz) and 284.D-5058 (PI: Rivinius). This work has made use of the VALD database (see \citealt{1999A&AS..138..119K}, and references therein), operated at Uppsala University, the Institute of Astronomy RAS in Moscow, and the University of Vienna. This work has used data from the European Space Agency (ESA) mission {\it Gaia} (\url{https://www.cosmos.esa.int/gaia}; \citealt{2016A&A...595A...1G,2021A&A...649A...1G}), processed by the {\it Gaia} Data Processing and Analysis Consortium (DPAC, \url{https://www.cosmos.esa.int/web/gaia/dpac/consortium}). DPAC funding has been provided by national institutions, in particular those participating in the {\it Gaia} Multilateral Agreement.

\section*{Data Availability}
Data from the \TESS~Mission is available on the Barbara A. Mikulski Archive for Space Telescopes (\url{mast.stsci.edu}). Resources supporting this work were provided by the NASA High-End Computing (HEC) Program through the NASA Advanced Supercomputing (NAS) Division at Ames Research Center for the production of SPOC data products. Data from SALT observations are made available to the public after a 6-month proprietary period. Both raw and processed spectra from XSHOOTER and UVES can be downloaded through the ESO Science Portal (\url{https://archive.eso.org/scienceportal/}). The ESPaDOnS data used in this article are available in the CFHT Science Archive at \url{https://www.cadc-ccda.hia-iha.nrc-cnrc.gc.ca/en/cfht/}.

\facilities{TESS, Gaia, HIPPARCOS, SALT (RSS spectropolarimeter), VLT:Kueyen (XSHOOTER spectrograph, UVES spectrograph), CFHT (ESPaDOnS spectrograph/spectropolarimeter)}

\software{SPOC \citep{jenkinsSPOC2016}, \texttt{astropy} \citep{astropy:2013, astropy:2018}, \texttt{numpy} \citep{harris2020array}, \texttt{matplotlib} \citep{Hunter:2007}, \texttt{scipy} \citep{2020SciPy-NMeth}, \texttt{pandas} \citep{reback2020pandas, mckinney-proc-scipy-2010}, \texttt{eleanor} \citep{2019ascl.soft04022F}, {\sc{starlink}} \citep{2014ASPC..485..391C}}

\bibliography{hd135348}{}
\bibliographystyle{aasjournal}


\label{lastpage}
\end{document}